\documentclass[fleqn,11pt]{article}
\usepackage[cp1251]{inputenc}
\usepackage{amsmath,amssymb}
\usepackage{xcolor}

\newtheorem{stat}{Statement }

\newcommand{\MI}{$\mathfrak{M}_1$}
\newcommand{\MII}{$\mathfrak{M}_2$}

\newcommand{\Fig}[3]{%
\begin{center}
\parbox{8cm}{%
\refstepcounter{figure}\includegraphics[width=8cm,height=#2cm]{#1} \noindent Fig. \thefigure:\quad
#3}\end{center}}

\usepackage{amsfonts,amssymb,cite}
\usepackage{graphicx}

\def\noi{\noindent}

\newcommand{\Title}[1]{\noi {{\Large\bf #1}}\\[1ex]}

\newcommand{\Author}[2]{\noi{\bf #1}\\[2ex]\noi{\normalsize\it #2}\\}

\newcommand{\Abstract}[1]{\vskip 2mm \begin{center}
        \parbox{16.4cm}{\small\noi #1} \end{center}\medskip}
\newcommand{\foom}[1]{\protect\footnotemark[#1]}

\def\nqq{\hspace*{-2em}}

\def\Jl#1#2{#1 {\bf #2},\ }

\def\ApJ#1 {\Jl{Astroph. J.}{#1}}
\def\CQG#1 {\Jl{Class. Quantum Grav.}{#1}}
\def\DAN#1 {\Jl{Dokl. AN SSSR}{#1}}
\def\GC#1 {\Jl{Grav. Cosmol.}{#1}}
\def\GRG#1 {\Jl{Gen. Rel. Grav.}{#1}}
\def\JETF#1 {\Jl{Zh. Eksp. Teor. Fiz.}{#1}}
\def\JETP#1 {\Jl{Sov. Phys. JETP}{#1}}
\def\JHEP#1 {\Jl{JHEP}{#1}}
\def\JMP#1 {\Jl{J. Math. Phys.}{#1}}
\def\NPB#1 {\Jl{Nucl. Phys. B}{#1}}
\def\NP#1 {\Jl{Nucl. Phys.}{#1}}
\def\PLA#1 {\Jl{Phys. Lett. A}{#1}}
\def\PLB#1 {\Jl{Phys. Lett. B}{#1}}
\def\PRD#1 {\Jl{Phys. Rev. D}{#1}}
\def\PRL#1 {\Jl{Phys. Rev. Lett.}{#1}}

\def\lal{&&\nqq {}}

\def\beq{\begin{equation}}
\def\eeq{\end{equation}}
\def\bear{\begin{eqnarray}}
\def\bearr{\begin{eqnarray} \lal}
\def\ear{\end{eqnarray}}
\def\earn{\nonumber \end{eqnarray}}

\def\e{{\,\rm e}}

\begin{document}
\thispagestyle{empty}
\twocolumn[

\vspace{1cm}

\Title{Gravitational - Scalar Instability of a Two-Component Degenerate System of Scalar Charged Fermions with Asymmetric Higgs Interaction \foom 1}

\Author{Yu. G. Ignat'ev$^{1,2}$}
    {$^1$Institute of Physics, Kazan Federal University, Kremlyovskaya str., 18, Kazan, 420008, Russia\\
    $^2$Lobachevsky Institute of Mathematics and Mechanics, Kazan Federal University, Kremlyovskaya str., 18, Kazan, 420008, Russia}

\Abstract
 {Based on the previously formulated mathematical model of a statistical system with scalar interaction of fermions and the theory of gravitational-scalar instability of a cosmological model based on a two-component statistical system of scalar-charged degenerate fermions, a numerical model of the cosmological evolution of gravitational-scalar perturbations is constructed and specific examples of the development of instability are given. Some features of the instability's development are investigated depending on the nature of the behavior of the unperturbed cosmological model. It is shown that unstable modes can appear at very early stages of cosmological expansion or contraction, and the duration of the unstable phase is comparable to tens of Planck scales. In this case, however, a very significant increase in unstable modes is possible due to the redistribution of energy between the components of the scalar doublet.
}
\bigskip

] 
\section*{Introduction}
In recent years, the existence of supermassive Black Holes with masses in the range $10^9\div10^{11} M_\odot$ at the centers of galaxies and quasars has been confirmed (see, for example, \cite{SMBH1}, \cite{SMBH2}). Currently, it is believed that supermassive Black Holes with masses of the order of $\sim10^9M_\odot $ are the central objects of luminous quasars observed at redshifts $ z> 6 $. At present, more than 200 quasars with $ z> 6 $ and several objects with $ z> 7 $ have been discovered. For example, a quasar with the largest redshift at $ z = 7.5 $, which corresponds to the age of the Universe of 650 million years, has an absolute luminosity of $ 1.4 \cdot10^{47} $ erg / sec and a central Black Hole mass of $ 1.6 \pm 0,4 \cdot10^9 M_\odot $ \cite{Fan}. The astrophysical origin of such supermassive Black Holes in the early Universe is still insufficiently understood, since observational data raise the question of the mechanism of formation and rapid growth of such objects in the early Universe.

The results of numerical simulations of \cite{Trakhtenbrot} impose a number of restrictions on the parameters of the formation of supermassive black holes. For example, it has been shown that light nuclei of Black Holes with a mass of $M\leqslant 10^3 M_\odot $, even with supercritical accretion, cannot grow to masses of the order of $10^8 M_\odot $ by $ z = 6 $. The formation of supermassive Black Holes with masses $10^8 \div 10^9 M_ \odot $ requires heavier nuclei $ M \sim 10^4 \div 10^6 M_\odot $ and gas-rich galaxies containing quasars. However, at present there are no sufficiently convincing models for the appearance of such heavy nuclei in the early Universe.

Interest in the mechanisms of formation of super\-mas\-sive Black Holes, taking into account the fact of their dominant presence in quasars, is caused, in particular, by the fact that such Black Holes are formed as components of quasars at rather early stages of the evolution of the Universe, before the formation of stars. This circumstance, in particular, opens up the possibility of the formation of supermassive Black Holes under conditions when scalar fields and baryonic dark matter can have a significant effect on this process. Note that the numerical simulation in \cite{Trakhtenbrot} was carried out within the framework of the standard gas acc\-re\-tion model, which does not take into account the possible influence of scalar fields on the formation of Black Holes. In this regard, we note the works \cite{Supermass_BH} - \cite{Soliton}, in which the possibility of the existence of \emph{scalar halos} and \emph{scalar hairs} in the neighbourhood of supermassive Black Holes is considered.

On the other hand, since the 80s, the Author has been carrying out targeted theoretical studies on the construction of a macroscopic theory of scalar charged particles (see, for example, \cite{Ignatev1} -- \cite{Yu_15GC}). In these models, elementary particles can have some charges with respect to each of the scalar fields, as a result of which the vacuum character of scalar fields is violated - in the right-hand sides of the corresponding field equations, densities of scalar charges appear. Initially, these studies pursued purely theoretical goals - the inclusion of interparticle scalar interactions in the scheme of the general relativistic kinetic theory. However, with the deve\-lop\-ment of the theory, the important role of scalar charged particles in cosmological models started to unravel. In particular, the class of such models was investigated by numerical simulation methods in the work \cite{ASS}, in which the possibility of controlling the stages of cosmological evolution using the parameters of scalarly charged fermions was shown. In \cite{Ignat_Sasha_G&G}, based on the study of degenerate systems of scalar-charged fermions with phantom interaction, it was suggested that Black Holes could form in the early Universe due to the \emph{gravitational-scalar instability} of a cosmological system consisting of degenerate systems of scalar-charged fermions \footnote{Further, to shorten the writing, the term ``degenerate statistical systems of fermions'' is replaced by ``degenerate fermions''.} In \cite{GC_21_1} -- \cite{GC_21_2} based on the developed theory of instability of a one-component system of scalarly charged degenerate fermions with the singlet Higgs scalar interaction in the so-called \emph{hard WKB-approximation} confirmed the assumption that such a system is unstable with respect to short-wavelength perturbations. These preliminary studies have shown the need for a comprehensive and deeper study of statistical systems of scalar charged particles. Further, speaking about the \emph{gravitational - scalar instability} of the cosmological system of scalar charged particles, we emphasize the fact that the growth mechanism of longitudinal perturbation modes is fundamentally different in the standard Jeans mechanism of gra\-vi\-ta\-tio\-nal instability and in the system of scalar charged fermions, primarily due to the peculiarities of interactions of scalar fields with scalar charged fermions. Let us recall that the main necessary condition for the development of gravitational Jeans instability is the condition of the nonrelativistic equation of state of matter (see, for example, \cite{Land_Field}). In the case of gravitational - scalar instability, this condition is removed.

Further, in \cite{TMF_Ign_Ign21}, two simplest models of interaction of fermions with an asymmetric scalar doublet\footnote{or an asymmetric Higgs scalar interaction (a pair of scalar fields, one of which is canonical, the second is phantom)} were proposed: in the first model, this interaction carried out by two types of different-sorted fermions, one of which is the source of the canonical scalar field, and the second is the phantom one (model \MI); in the second model, there is one kind of fermions with a pair of charges - canonical and phantom (model \MII). A qualitative analysis of the dynamic system of the \MI\ model was also carried out there. In the work \cite{Ignat_GC21}, a numerical integration of the equations of the cosmological model \MI\ was carried out, on the basis of which the features of this model were revealed, in particular, the possibility of the existence of cosmological contraction phases, oscillations of the Hubble parameter, as well as the possibility of universes with a finite lifetime.

The purpose of this work is, firstly, to construct a numerical model of the development of the in\-sta\-bi\-lity of perturbations in the model \MI\ of a two-component statistical system with asymmetric scalar interaction of fermions, removing the condition of the hard WKB-approximation and, secondly, a preliminary study of this model in order to identify the most significant features of the development of instability. In particular, our main task was to confirm the fundamental possibility of instability of longitudinal perturbations of the cosmological model in the early Universe. At the same time, we deliberately did not select the possible values of the model parameters, assuming all of them to be equal to 1 (Section 3) in order to show the generality of the picture of the development of early instability. The specific parameters of the model: the masses of fermions, their scalar charges, etc., are practically unknown to us. In the next work, we plan to conduct a more detailed study of the development of instability depending on the parameters of the model in order to identify the possibility of early formation of supermassive Black Holes.

\section{Two-Component Cosmological\newline System of Degenerate Scalar - Charged Fermions}
\subsection{General relations for a self - gravita-\newline ting system of degenerate scalar - charged fermions}
Consider a cosmological model based on a self-gravitating two-component system of scalar singly charged degenerate fermions interacting through a pair of scalar Higgs fields, canonical, $\Phi$, and phantom, $\varphi$. This model is described by the Einstein system of equations
\begin{equation}\label{Eq_Einst_G}
R^i_k-\frac{1}{2}\delta^i_k R=8\pi T^i_k+ \delta^i_k \Lambda_0,
\end{equation}
where
\[T^i_k=T^i_{(s)k}+T^i_{(p)k},\]
and $T^i_{(s)k}$ -- the energy-momentum tensor of scalar fields, $T^i_{(p) k} $ is the energy-momentum tensor of particles, and $\Lambda_0$ is the seed value of the cosmological constant associated with its observed value $\Lambda$ by the relation:
\begin{equation}\label{lambda0->Lambda}
\Lambda=\Lambda_0-\frac{1}{4}\sum\limits_r \frac{m^4_r}{\alpha_r}.
\end{equation}
Further, $L_s$ is the Lagrange function of noninteracting scalar fields of canonical ($\Phi_c \ equiv \Phi$) and phantom ($\Phi_f\equiv\varphi $)
\begin{eqnarray} \label{Ls}
L_s=\frac{1}{16\pi}\sum\limits_r(\epsilon_r g^{ik} \Phi_{r,i} \Phi _{r,k} -2V_r(\Phi_r)),
\end{eqnarray}
\begin{eqnarray}
\label{Higgs}
V_r(\Phi_r)=-\frac{\alpha_r}{4} \left(\Phi_r^{2} -\frac{m_r^{2} }{\alpha_m}\right)^{2}
\end{eqnarray}
are the potential energy of the corresponding scalar fields, $\alpha_c \equiv \alpha$ and $\alpha_f \equiv \beta$ are their self-action constants, $ m_c \equiv m$ and $m_f \equiv \mathfrak {m} $ - their masses of quanta, $\epsilon_r = \pm 1$ are indicators (the `` + '' sign corresponds to canonical scalar fields, the `` - '' sign - to phantom fields). As a carrier of scalar charges, we consider a two-component degenerate system of fermions, in which the carriers of the canonical charge of the $z$-fermions have the canonical charge $ e_z $ and the Fermi momentum $\pi_{(z)}$, and the carriers of the phantom charge of the $\zeta $ -fermions have phantom charge $e_\zeta$ and Fermi momentum $\pi_{(\zeta)}$. The dynamic masses of these fermions in the case of zero seed masses are \cite{TMF_Ign_Ign21}
\begin{equation}\label{m_*(pm)}
m_{z}=e_z\Phi;\qquad m_{\zeta}=e_\zeta\varphi.
\end{equation}
Further, the tensor of energy - momentum of scalar fields with respect to the Lagrange function \eqref{Ls} is:
\begin{eqnarray}\label{T_s}
\!\!T^i_{(s)k}=\!\frac{1}{16\pi }\sum\limits_r\bigl(2\e_r\Phi^{,i}_{r} \Phi _{r,k} -e_r\delta^i_k\Phi _{r,j} \Phi _r^{,j}
 \nonumber\\+2V_r(\Phi_r)\delta^i_k \bigr),
\end{eqnarray}
and the tensor of energy - momentum of an equi\-li\-b\-rium statistical system is equal to:
\begin{equation}\label{T_p}
T^i_{(p)k}=(\varepsilon_p+p_p)u^i u_k-\delta^i_k p_p,
\end{equation}
where $ u^i$ is a vector of the macroscopic velocity of the statistical system, $\varepsilon_p$ and $p_p$ are its energy density and pressure.

\emph{Macroscopic scalars}, i.e., scalar functions that determine the macroscopic characteristics of the statistical system \footnote{its energy density, pressure, scalar charge density, etc.} for a two-component statistical system of degenerate fermions are equal (see, for example, \cite{ Yu_15GC}, \cite{TMF_Ign_Ign21}):
\begin{eqnarray}
\label{2_3a_2}
\varepsilon_p=\frac{e^4_z \Phi^4}{8\pi^2}F_2(\psi_z)+\frac{e^4_\zeta \varphi^4}{8\pi^2}F_2(\psi_\zeta);\\
\label{2_3b_2}
p_p  =\displaystyle \frac{e^4_z \Phi^4}{24\pi^2}(F_2(\psi_z)-4F_1(\psi_z))+\nonumber\\
\frac{e^4_\zeta \varphi^4}{24\pi^2}(F_2(\psi_\zeta)-4F_1(\psi_\zeta));\\
\label{2_3c}
\displaystyle
\sigma^z=\frac{e_z^4 \Phi^3}{2\pi^2}F_1(\psi_z);\qquad \sigma^\zeta=\frac{e_\zeta^4 \varphi^3}{2\pi^2}F_1(\psi_\zeta),
\end{eqnarray}
where $\sigma^z$ and $\sigma^\zeta$ are the densities of scalar charges $e_z$ and $e_\zeta$ and
\begin{equation}\label{psi_zzeta}
\psi_z=\frac{\pi_{(z)}}{|e_z\Phi|}; \qquad \psi_\zeta=\frac{\pi_{(\zeta)}}{|e_\zeta\varphi|}.
\end{equation}
To shorten the writing, the functions $F_1(\ psi)$ and $F_2(\psi)$ have been introduced:
\begin{equation}\label{F_1}
F_1(\psi)=\psi\sqrt{1+\psi^2}-\ln(\psi+\sqrt{1+\psi^2});
\end{equation}
\begin{equation}
\label{F_2}
F_2(\psi)=\psi\sqrt{1+\psi^2}(1+2\psi^2)-\ln(\psi+\sqrt{1+\psi^2}).
\end{equation}
Finally, the equations of scalar fields for the investigated system take the form:
\begin{eqnarray}\label{Box(Phi)=sigma_z}
\Box \Phi + m^2\Phi-\alpha\Phi^3 = -\frac{4}{\pi^2}e^4_z\Phi^4 F_1(\psi_z),\\
\label{Box(varphi)=sigma_zeta}
-\Box \varphi + \mathfrak{m}^2\varphi-\beta\varphi^3 = -\frac{4}{\pi^2}e^4_\zeta\varphi^4 F_1(\psi_\zeta).
\end{eqnarray}

Note that the presented self-consistent system of macroscopic equations, including the above expres\-sions for macroscopic scalars \eqref{2_3a_2} -- \eqref{psi_zzeta}, are strict con\-se\-qu\-ences of the general relativistic kinetic theory, namely, the transport equations, for the case of thermodynamic equilibrium.
\subsection{Unperturbed State of a Homogeneous and Isotropic Cosmological System}

Let us further consider the spatially flat model of the Friedman universe \footnote{We use a metric with the signature $ (--- +) $, the Ricci tensor is determined by the convolution of the first and third indices of the curvature tensor (see, for example, \cite{Land_Field})}:
\begin{eqnarray}\label{ds0}
ds_0^2=dt^2-a^2(t)(dx^2+dy^2+dz^2)\equiv\nonumber\\
a^2(\eta)(d\eta^2-dx^2-dy^2-dz^2),\quad(t=\int ad\eta).
\end{eqnarray}
It can be proved that a strict consequence of the general relativistic kinetic theory for statistical systems of completely degenerate fermions is the Fermi momentum conservation law $\pi_{(a)}$ of each component \footnote{For a one-component system of degenerate fermions see \cite{Ignat14_2}, and for a multicomponent see \cite{TMF_Ign_Ign21}.}
 \begin{equation}\label{ap}
 a(t)\pi_{(a)}(t)=\mathrm{Const}.
 \end{equation}
Further, for definiteness, we set $a(0)=1$ and
\begin{eqnarray}\label{a-xi}
\xi=\ln a;\quad \xi\in(-\infty,+\infty); \quad \xi(0)=0,\\
\label{psi0}
\pi_{(z)}=\pi_c \mathrm{e}^{-\xi},\; \pi_{(\zeta)}=\pi_f \mathrm{e}^{-\xi},\nonumber\\
 (\pi_c=\pi_{(z)}(0),\pi_f=\pi_{(\zeta)}(0)),
\end{eqnarray}
Let us write down the complete normal system of Einstein equations and scalar fields $\Phi (t)$ and $\varphi(t)$ for this two-component system of scalarly charged degenerate fermions \cite{TMF_Ign_Ign21} \footnote{For a scalar singlet, this system was obtained in \cite{Ignat20_1}}.
In an explicitly nonsingular form, the normal system of ordinary differential equations of the investigated model has the form ($\xi \equiv \ln a $, $ \dot{f} \equiv df / dt $):
\begin{eqnarray}
\label{dxi/dt-dPhi_Phi}
\dot{\xi}=H;\qquad \dot{\Phi}=Z;\qquad \dot{\varphi}=z;\\
\label{dH/dt_M1}
\dot{H}=-\frac{Z^2}{2}+\frac{z^2}{2}-\frac{4\mathrm{e}^{-3\xi}}{3\pi}\times\nonumber\\
\biggl(\pi_c^3\sqrt{\pi_c^2\mathrm{e}^{-2\xi}+e^2\Phi^2}+\pi_f^3\sqrt{\pi_f^2\mathrm{e}^{-2\xi}+\epsilon^2\varphi^2}\biggr);
\end{eqnarray}
\begin{eqnarray}
\label{dZ/dt_M1}
\dot{Z}=-3HZ-m^2\Phi+\alpha\Phi^3-\nonumber\\[6pt]
\frac{4e^2\pi_c\mathrm{e}^{-\xi}}{\pi}\Phi\sqrt{\pi^2_c \mathrm{e}^{-2\xi}+e^2\Phi^2}+\nonumber\\
\frac{4e^4}{\pi}\Phi^3\ln\biggl(\frac{\pi_c\mathrm{e}^{-\xi}+\sqrt{\pi^2_c \mathrm{e}^{-2\xi}+e^2\Phi^2}}{|e\Phi|} \biggr);
\end{eqnarray}
\begin{eqnarray}
\label{dzZ/dt_M1}
\dot{z}=-3Hz+\mathfrak{m}^2\varphi-\beta\varphi^3+\nonumber\\[6pt]
\frac{4\epsilon^2\pi_f\mathrm{e}^{-\xi}}{\pi}\varphi\sqrt{\pi^2_f \mathrm{e}^{-2\xi}+\epsilon^2\varphi^2}-\nonumber\\
\frac{4\epsilon^4}{\pi}\varphi^3\ln\biggl(\frac{\pi_f\mathrm{e}^{-\xi}+\sqrt{\pi^2_f \mathrm{e}^{-2\xi}+\epsilon^2\varphi^2}}{|\epsilon\varphi|} \biggr).
\end{eqnarray}
The system of equations \eqref{dxi/dt-dPhi_Phi} -- \eqref{dzZ/dt_M1} as its first integral has the total energy integral \cite{TMF_Ign_Ign21}, which can be used to determine the initial value of the function $H(t)$
\begin{eqnarray}
\frac{Z^2}{2}+\frac{z^2}{2}-\frac{m^2\Phi^2}{2}+\frac{\alpha\Phi^4}{4}-\frac{\mathfrak{m}^2\varphi^2}{2}+\frac{\beta\varphi^4}{4} \nonumber\\
-\frac{e^{-\xi}}{\pi}\biggl(\pi_c\sqrt{\pi_c^2\mathrm{e}^{-2\xi}+e^2\Phi^2}\bigl(2\pi^2_c\mathrm{e}^{-2\xi}+e^2\Phi^2\bigr)\nonumber\\
+\pi_f\sqrt{\pi_f^2\mathrm{e}^{-2\xi}+\epsilon^2\varphi^2}\bigl(2\pi^2_f \mathrm{e}^{-2\xi}+\epsilon^2\varphi^2\bigr)\biggr)\nonumber
\end{eqnarray}
\begin{eqnarray}\label{SurfEinst_M1}
+\frac{e^4\Phi^4}{\pi}\ln\biggl(\frac{\pi_c\mathrm{e}^{-\xi}+\sqrt{\pi^2_c \mathrm{e}^{-2\xi}+e^2\Phi^2}}{|e\Phi|} \biggr)\nonumber\\
+\frac{\epsilon^4\varphi^4}{\pi}\ln\biggl(\frac{\pi_f\mathrm{e}^{-\xi}+\sqrt{\pi^2_f \mathrm{e}^{-2\xi}+\epsilon^2\varphi^2}}{|\epsilon\varphi|} \biggr)\nonumber\\
+3H^2-\Lambda=0.
\end{eqnarray}
The \eqref{dxi/dt-dPhi_Phi} -- \eqref{SurfEinst_M1} equations describe the \MI\ cosmo\-lo\-gical model proposed in \cite{TMF_Ign_Ign21} and explored in \cite{Ignat_GC21}.
\section{Results of the Theory of\newline Gravitational - Scalar Instability of the \MI\ Model in the Shortwave Limit}
Let us briefly describe the main results of the work \cite{STFI_20}, in which the evolution of gravitational - scalar perturbations of the metric \eqref{ds0} and scalar fields in the \MI \ model for the case of purely longitudinal perturbations of the metrics and \eqref{ds0} is investigated, representing the perturbed metric in \cite{Land_Field} (see \cite{Lifshitz} for details)
\begin{eqnarray}
\label{metric_pert}
ds^2=ds^2_0-a^2(\eta)h_{\alpha\beta}dx^\alpha dx^\beta,
\end{eqnarray}
where $ ds_0 $ is the unperturbed spatially flat Fri\-ed\-mann metric \eqref{ds0} in conformally flat form
and for defi\-ni\-teness, the wave vector is directed along the $Oz$ axis:
\begin{eqnarray}\label{nz1}
 h_{11}=h_{22} =\frac{1}{3}[\lambda(t)+\frac{1}{3}\mu(t)]\mathrm{e}^{inz};\nonumber\\
\label{nz13}
h=\mu(t)\mathrm{e}^{inz};\; h_{12}=h_{13}= h_{23}=0;\nonumber\\
\label{nz2}
h_{33}=\frac{1}{3}[-2\lambda(t)+\mu(t)]\mathrm{e}^{inz}.
\end{eqnarray}
In this case, the matter in the \MI\ model is completely determined by four scalar functions - $\Phi(z,\eta)$, $\varphi(z,\eta)$, $\pi_{(z)} (z, \eta) $ and $\pi_{(\zeta)} (z, \eta) $ and the velocity vector $ u^i (z, \eta) $. Let us expand these functions in a series in terms of the smallness of the perturbations with respect to the corresponding functions against the background of the Friedmann metric \eqref{ds0} \footnote{For scalar singlets, see \cite{GC_21_1}. To avoid cumbersome notation, we have retained the notation for the perturbed values of functions, distinguishing them only by arguments.}:
\begin{eqnarray}\label{dF-drho-du}
\Phi(z,\eta)=\Phi(\eta)+\delta\Phi(\eta)\mathrm{e}^{inz};\nonumber\\
\varphi(z,\eta)=\varphi(\eta)+\delta\varphi(\eta)\mathrm{e}^{inz};\nonumber\\
\pi_{(z)}(z,t)=\pi_{(z)}(\eta)(1+\delta(\eta)\mathrm{e}^{inz});\nonumber\\
\pi_{(\zeta)}(z,t)=\pi_{(\zeta)}(\eta)(1+\delta(\eta)\mathrm{e}^{inz});\nonumber\\
\sigma^z(z,\eta)= \sigma^z(\eta)+\delta\sigma^z(\eta)\mathrm{e}^{inz};\\
\sigma^\zeta(z,\eta)= \sigma^\zeta(\eta)+\delta\sigma^\zeta(\eta)\mathrm{e}^{inz};\nonumber\\
u^i=\frac{1}{a}\delta^i_4+\delta^i_3 v(\eta)\mathrm{e}^{inz},\nonumber
\end{eqnarray}
where $\delta \Phi (\eta) $, $ \delta \varphi (\eta) $, $ \delta (\eta) $, $ s_z (\eta), s_\zeta (\eta) $ and $ v (\eta) $ are functions of the first order of smallness in comparison with their unperturbed values. Note that the relative perturbations of the Fermi mo\-men\-tum $\delta (\eta)$ do not depend on the type of fermions \cite{STFI_20}.

In \cite{STFI_20}, the evolution of longitudinal gravitational scalar perturbations of the \MI \ model is investigated in the short-wave and small-charge approximation:
\begin{eqnarray}\label{WKB}
n\eta\gg 1;\; n^2\gg  e^4_{(a)},\; a^2\{m^2,\mathfrak{m}^2\}\gg e^4_{(a)}.
\end{eqnarray}
At the same time, in contrast to the works \cite{GC_21_1} -- \cite{GC_21_2}, the condition of the hard WKB approximation was not imposed in this paper, which allows considering also sufficiently long wavelengths:
\[n^2\gtrsim a^2\{ m^2\Phi,\alpha\Phi^3,\mathfrak{m}^2\varphi,\beta\varphi^3\}.\]
In accordance with the WKB method, we represent the perturbation functions $ f (\eta) $ in the form
\begin{equation}\label{Eiconal}
f=\tilde{f}(\eta) \cdot \mathrm{e}^{i\int u(\eta)d\eta}; \quad (|u|\sim n \gg 1),
\end{equation}
where $\tilde{f} (\eta)$ and $u(\eta)$ are functions of the amplitude and eikonal of the perturbation that vary weakly along with the scale factor.

Substitution of perturbations \eqref{dF-drho-du} in the form \eqref{Eiconal} into the equations of scalar and gravitational fields of the first order of the perturbation theory in the WKB approximation \eqref{Eiconal}\footnote {for some details see, for example, \cite{GC_21_1}, \cite{GC_21_2}} results in the following \cite{STFI_20}.

Let us introduce the following notation:
\begin{eqnarray}\label{de_delta}
 \varepsilon_p^\delta=\frac{e_z^4\Phi^4\psi_z^3\sqrt{1+\psi_z^2}
+e_\zeta^4\varphi^4\psi_\zeta^3\sqrt{1+\psi_\zeta^2}}{\pi^2} ;\nonumber\\
\varepsilon_p^\Phi=\frac{e_z^4\Phi^3}{2\pi^2}F_1(\psi_z);
\varepsilon_p^\varphi=\frac{e_\zeta^4\varphi^3}{2\pi^2}F_1(\psi_\zeta);\\
\Delta_\Phi=\frac{\varepsilon_p^\Phi}{ 8\pi \varepsilon_p^\delta};\;
\Delta_\varphi=\frac{\varepsilon_p^\varphi}{ 8\pi \varepsilon_p^\delta};\nonumber
\end{eqnarray}
\begin{eqnarray}
\!\!\!S^z_\Phi=\frac{e^4_z\Phi^2}{2\pi^2}\!\!\biggl(3F_1(\psi_z)-\frac{\psi^3_z}{\sqrt{1+\psi^2_z}}-\frac{\psi^2_z}{\sqrt{1+\psi^2_z}}\Delta_\Phi\biggr)
;\nonumber
\end{eqnarray}
\begin{eqnarray}
S^z_\varphi=-\frac{e^4_z\Phi^2\psi^2_z}{2\pi^2\sqrt{1+\psi^2_z}}\Delta_\varphi; S^\zeta_\Phi=-\frac{e^4_\zeta\varphi^2\psi^2_\zeta}{2\pi^2\sqrt{1+\psi^2_\zeta}}\Delta_\Phi; \nonumber
\end{eqnarray}
\begin{eqnarray}
\label{s_zeta-delta}
\!\!\!\!\! S^\zeta_\varphi=\frac{e^4_\zeta\varphi^2}{2\pi^2}\!\!\biggl(3F_1(\psi_\zeta)-\frac{\psi^3_\zeta}{\sqrt{1+\psi^2_\zeta}}-
\frac{\psi^2_\zeta}{\sqrt{1+\psi^2_\zeta}}\Delta_\varphi\biggr);\nonumber
\end{eqnarray}
\begin{eqnarray}
\!\!\!p_p^\delta=\frac{1}{\pi^2}\biggl(\frac{e_z^4\Phi^4\psi_z^4}{\sqrt{1+\psi_z^2}}
+\frac{e_\zeta^4\varphi^4\psi_\zeta^4}{\sqrt{1+\psi_\zeta^2}}\biggl)>0\hskip 5mm;\nonumber\end{eqnarray}
\begin{eqnarray}\label{P^Phi}
\!\!\!P^\Phi=\frac{e^4_z\Phi^3}{2\pi^2}F_1(\psi_z)-p_p^\delta\Delta_\Phi; \nonumber\\
P^\varphi=\frac{e^4_\zeta\varphi^3}{2\pi^2}F_1(\psi_\zeta)-p_p^\delta\Delta_\varphi,
\end{eqnarray}
with the help of which we find the following expres\-sions for macroscopic scalars:
\begin{eqnarray}\label{s_z-delta}
\!\!\!\delta\sigma^z=\frac{e^4_z\Phi^3\psi^2_z}{48\pi^3 a^2\varepsilon^\delta_p\sqrt{1+\psi^2_z}}n^2\nu +S^z_\Phi\delta\Phi+S^z_\varphi\delta\varphi;&\nonumber\\
\!\!\!\delta\sigma^\zeta=\frac{e^4_\zeta\varphi^3\psi^2_z}{48\pi^3 a^2\varepsilon^\delta_p\sqrt{1+\psi^2_\zeta}}n^2\nu +S^\zeta_\Phi\delta\Phi+S^\zeta_\varphi\delta\varphi;&\nonumber\\
\label{dp-delta}
\delta p_p=\frac{p^\delta_p n^2}{24\pi^3a^2\varepsilon^\delta_p}\nu +P^\Phi\delta\Phi+P^\varphi\delta\varphi. &\nonumber
\end{eqnarray}

We also introduce the coefficients $\gamma_{\alpha\beta}$ of the \emph{dispersion relation}
\begin{eqnarray}
\label{gamma_ik}
\gamma_{11}\equiv-a^2(m^2-3\alpha\Phi^2+8\pi S^z_\Phi);\nonumber\\
\gamma_{22}\equiv a^2(\mathfrak{m}^2-3\beta\varphi^2+8\pi S^\zeta_\varphi);\nonumber\\
\gamma_{33}\equiv \frac{1}{3}+\frac{p^\delta_p}{\varepsilon^\delta_p};\;
\gamma_{12}\equiv -8\pi a^2S^z_\varphi;\nonumber\\
 \gamma_{21}\equiv 8\pi a^2S^\zeta_\Phi;\;
\gamma_{31}\equiv -3a^2[\Phi(m^2-\alpha\Phi^2)-8\pi P^\Phi]; \nonumber\\
\gamma_{32}\equiv  -3a^2[\varphi(\mathfrak{m}^2-\beta\varphi^2)-8\pi P^\varphi]; \nonumber\\
 \gamma_{13}\equiv \frac{e^4_z\Phi^3\psi^2_z}{6\pi^2\varepsilon^\delta_p\sqrt{1+\psi^2_z}};\; \gamma_{23}\equiv-\frac{e^4_\zeta\varphi^3\psi^2_\zeta}{6\pi^2\varepsilon^\delta_p\sqrt{1+\psi^2_\zeta}}. \nonumber
\end{eqnarray}

Let us now formulate the following statements of the work \cite{STFI_20}.

\begin{stat}\label{u_123}
Under the WKB conditions \eqref{WKB}, the equations for the perturbations \eqref{dF-drho-du} have the following solutions for the eikonal function:
\end{stat}
\begin{eqnarray}\label{u_3-u_1}
u^\pm_{(0)}=0; \quad u^\pm_{(1)}=\pm n\ \sqrt{\frac{1}{3}+\frac{p^\delta_p}{\varepsilon^\delta_p}};\\
\label{u_1_pm}
u^\pm_{(2\pm)}=\pm\biggl[n^2-\frac{1}{2}(\gamma_{11}+\gamma_{22})\times\nonumber\\
\pm\frac{1}{2}\sqrt{(\gamma_{11}+\gamma_{22})^2+4\gamma_{12}\gamma_{21}}\biggr]^{\frac{1}{2}},
\end{eqnarray}
\emph{where the upper signs correspond to the signs before the outer radical, the lower ones correspond to the signs before the inner one.}
In this case, the zero modes $u^\pm_{(0)} = 0$ correspond to per\-tur\-ba\-tions of the metric $\lambda$ and are eliminated by admissible transformations of the metric. The modes $u^\pm_{(1)}$ correspond to per\-tur\-ba\-tions of the metric $\mu$, and the modes $u^\pm_{(2\pm)}$ correspond to per\-tur\-ba\-tions of the scalar fields $\delta\Phi$ and $\delta\varphi$.

Next, we introduce the functions
\begin{eqnarray}\label{Gamma}
\Gamma(n,\eta)\equiv n^4-n^2(\gamma_{11}+\gamma_{22})\nonumber\\+\gamma_{11}\gamma_{22}-\gamma_{12}\gamma_{21},\\
\label{Upsilon}
\Upsilon(n,\eta) \equiv 2n^2-(\gamma_{11}+\gamma_{22}).
\end{eqnarray}
Then:
\begin{stat}\label{regions}
\end{stat}

\noindent\textbf{1.}\emph{ When conditions are met simultaneously }
\begin{eqnarray}
\Gamma>0,\Upsilon>0
\end{eqnarray}
\emph{the solutions $u^\pm_{(2\pm)}$ -- real or complex conjugate;}\\
\noindent \textbf{2.} \emph{When conditions are met simultaneously }
\begin{eqnarray}
\Gamma>0,\Upsilon<0
\end{eqnarray}
\emph{the solutions $u^\pm_{(2\pm)}$ -- conjugate;}\\
\noindent \textbf{3.}\emph{ If the condition is met }
\begin{eqnarray}
\Gamma<0
\end{eqnarray}
\emph{the solution $u^\pm_{(2+)}$ is real, and $u^\pm_{(2-)}$ is imaginary.}

Due to the linearity of the perturbations, the final expressions for the perturbations according to the formula \eqref{Eiconal} and the found values of the eikonal
functions \eqref{u_3-u_1} -- \eqref{u_1_pm} can be written in the form:
\begin{eqnarray}\label{Sol_perts}
f=\mathrm{e}^{inz}\sum\limits_\pm\tilde{f}_{(1)}^{\pm}(\eta)\mathrm{e}^{i\int u^\pm_{(1)}(\eta)d\eta}+\nonumber\\
\mathrm{e}^{inz}\sum\limits_\pm\sum\limits_\pm\tilde{f}^{\pm}_{(2)\pm}(\eta)\mathrm{e}^{i\int u^\pm_{(2)\pm}(\eta)d\eta},
\end{eqnarray}
where $\tilde{f}_{(1)}^{\pm}(\eta)$ and $\tilde{f}^{\pm}_{(2)\pm} (\eta)$ - slowly varying amplitudes of perturbations $\mu(\eta, z) $, $ \delta\Phi (\eta, z)$, $\delta\varphi (\eta, z)$, corresponding to the above modes of oscillations $u^\pm_{(1)}, u^+_{(2 \pm)}$. Since according to \eqref{de_delta} and \eqref{dp-delta} $\varepsilon_p^delta>0$ $p_p^\delta>0$, then $u^\pm_{(1)} $ are real functions, therefore, they correspond to retarded and advanced waves with weakly varying amplitudes. Thus, according to the statement \ref{u_123}, unstable modes can only correspond to the eikonal functions $u^\pm_{(2) \pm} (\eta) $ in the range of their complex values.

Let
\begin{equation}\label{damp}
u^\pm_{(2)\pm}(\eta)=\omega(n,\eta)\pm i\vartheta(n,\eta),
\end{equation}
where $\vartheta(n,\eta) \geqslant0 $, then these values of the eikonal function correspond to damped and increasing oscil\-lation modes with amplitudes:
\begin{equation}\label{instability}
\tilde{f}^-(\eta)\mathrm{e}^{ -\int\vartheta(n,\eta)d\eta},\; \tilde{f}^+(\eta)\mathrm{e}^{+\int\vartheta(n,\eta)d\eta}.
\end{equation}
The increasing oscillation mode $\tilde{f}^+$ exactly corres\-ponds to the instability of the homogeneous unper\-turbed state of the cosmological model. As we noted above, this mode corre\-sponds to perturbations of scalar fields. Therefore, the instability has an essen\-tially gravitational-scalar nature. We will return to this issue below.

According to the statements \ref{u_123} -- \ref{regions} and the formulas \eqref{gamma_ik}, the investigated cosmological system is unstable in the following regions:
\begin{eqnarray}\label{O1}
\Omega_1(n,\eta): \; \{\Gamma(n,\eta)>0,
\gamma_{11}+\gamma_{22}>0\};\nonumber\\
n^2_-<n^2<n^2_+;
\end{eqnarray}
\begin{eqnarray}\label{O2}
\Omega_2(n,\eta): \; \left\{\begin{array}{l}
\{(\gamma_{11}-\gamma_{22})^2+4\gamma_{12}\gamma_{21}>0, \\
\gamma_{11}\gamma_{22}-\gamma_{12}\gamma_{21}>0, \\
\gamma_{11}+\gamma_{22}>0\},\\
 n^2<\frac{1}{2}(\gamma_{11}+\gamma_{22});\\[3pt]
 \hline\\[-3pt]
\{(\gamma_{11}-\gamma_{22})^2+4\gamma_{12}\gamma_{21}<0,   \\
\gamma_{11}+\gamma_{22}>0\},\\
\{n^2<n^2_-\}\cup\{ n^2>n^2_+\}.\\
\end{array}\right.
\end{eqnarray}
where
\begin{eqnarray}\label{n^2}
\!\!\!\!n^2_\pm=\frac{1}{2}\biggl[\gamma_{11}+\gamma_{22}\pm\sqrt{(\gamma_{11}-\gamma_{22})^2+4\gamma_{12}\gamma_{21}}.\biggr].
\end{eqnarray}
\section{Numerical Simulation of the\newline Cosmological Evolution of\newline Perturbations in the \MI Model}
In \cite {STFI_20} examples of instability regions of the investigated cosmological system were given, under the assumption that the functions $ a (t) $, $ \Phi (t) $ and $\varphi(t)$ are given. For the instabilities described above to arise, it is necessary to realize the required values of these parameters in the process of cosmological evolution. Clarification of this issue, in turn, requires a joint numerical simulation of the evolution of the unperturbed cosmological model and its perturbations.
\subsection{Evolution of the Local Increment\newline of Oscillation Increase}
To find the unperturbed functions $a(t)$, $\Phi (t)$ and $\varphi (t)$, it is necessary to determine the numerical solution of the system of equations \eqref{dxi/dt-dPhi_Phi} -- \eqref{dzZ/dt_M1} with the integral condition \eqref{SurfEinst_M1}. This problem was solved in the work \cite{Ignat_GC21}.
The solution to this problem will allow us to determine the macroscopic scalars \eqref{de_delta}- - \eqref{dp-delta}, then the matrix elements $\gamma_{\alpha\beta}(t)$ and, ultimately, the eikonal functions $u^\pm_{(2\pm)} (t) $ \eqref{u_1_pm}. The problem is that the unperturbed solution is determined using the cosmological time $t$, while the perturbations \eqref{Sol_perts} are determined by the time variable $\eta$. Passing to cosmological time according to the formula \eqref{ds0} in expressions \eqref{Sol_perts}, we obtain according to \eqref{damp}
\begin{eqnarray}
i\int\limits_{\eta_0}^\eta u^\pm_{(2)\pm}d\eta=i\int\limits_0^t \frac{u^\pm_{(2)\pm}(t)}{a(t)}dt\nonumber\\
\equiv i\int\limits_0^t \omega^\pm_\pm dt-\int\limits_0^t \gamma^\pm_\pm dt,
\end{eqnarray}
where $\omega (t)$ and $\gamma (t)$ are the local frequency and decrement /increment of damping /increasing of oscillations:
\begin{equation}\label{g,o}
\!\!\!\omega^\pm_\pm=\mathrm{e}^{-\xi(t)}\mathrm{Re}(u^\pm_{(2)\pm});\; \gamma^\pm_\pm=\mathrm{e}^{-\xi(t)}\mathrm{Im}(u^\pm_{(2)\pm}).
\end{equation}

Moving on to numerical modeling, further, to shorten the letter, we will set a set of fundamental parameters of the \MI\ model using an ordered list (see \cite{Ignat_GC21})
\[\mathbf{P}=[[\alpha,m,e,m_c,\pi_c],[\beta,\mu,\epsilon,m_f,\pi_f],\Lambda]\]
and set the initial conditions as an ordered list
\[\mathbf{I}=[\Phi_0,Z_0,\varphi_0,z_0,\kappa],\]
where $\kappa = \pm1$, and the value $\ kappa = + 1$ corres\-ponds to the non-negative initial value of the Hubble parameter $H_0 = H _ + \geqslant0$, and the value $\kappa= -1$ corresponds to the negative initial value of the Hubble parameter $H_0 =H_- <0 $. In this case, using the autonomy of the dynamical system, we everywhere set $\xi(0) = 0$. Thus, the \MI\ model is determined by 11 fundamental parameters and 5 initial conditions. The large number of model parameters significantly increases the complexity of the model study.

\subsection{Standard Model Behavior}
Consider first the \MI\ model with ``standard behavior'', in which $\xi(t)$ is a monotonically increasing function and $H\geqslant0$:
\begin{eqnarray}\label{P1}
\mathbf{P_1}=[[1,1,1,0.1],[1,1,1,0.1],0.1],\nonumber\\
\mathbf{I_1}=[1,0,0.00001,0,1].
\end{eqnarray}
 Fig. \ref{ignatev1} and \ref{ignatev2} shows the evolution of the scale function $\xi (t)\equiv\ln a(t)$ and the Hubble parameter $H (t) \equiv \dot{\ xi}$ for this case.
\Fig{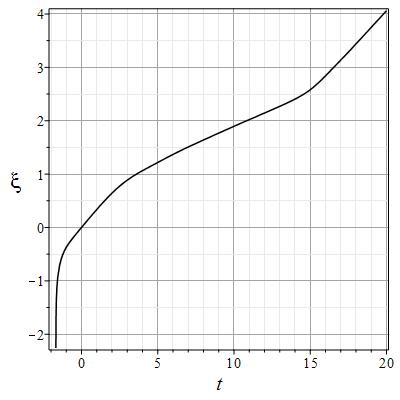}{8}{\label{ignatev1}Evolution of the scale function $\xi(t)$ for model parameters \MI\  $\mathbf{P_1}$ and initial conditions $\mathbf{I_1}$ \eqref{P1}.}
\Fig{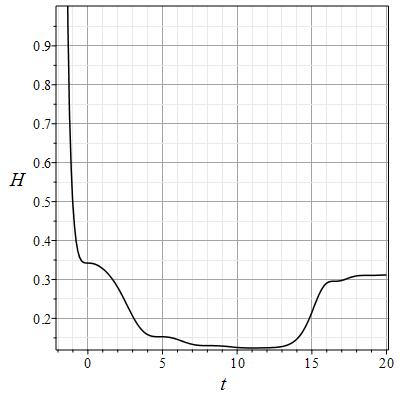}{8}{\label{ignatev2}Evolution of the Hubble parameter $H(t)$ for model parameters \MI\ $\mathbf{P_1}$ and initial conditions $\mathbf {I_1} $ \eqref{P1}.}
Further, Fig. \ref{ignatev3} shows the cosmological evolution of the local oscillation frequency $\omega^\pm(t)$ and the oscillation increasing increment $\gamma^\pm(t)$ corresponding to the modes $u^\pm_{(2) -}$ for the wave number $n =5$. It can be seen in this figure that in the region in which there are growing and damping modes, the oscillations of the system stop, thus, in this region, standing increase and damping modes arise. On the contrary, in the region where the local oscillation frequency is not equal to zero, there is no increase / damping of oscillations at all, excluding, of course, the standard geometric drop in amplitude due to expansion.
\Fig{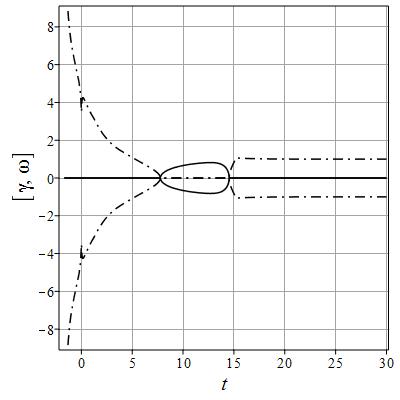}{8}{\label{ignatev3}Evolution of the local frequency $\omega^\pm(t) $ -- dash - dotted line and increment / decrement of increase / damping of the oscillation amplitude $\gamma^\pm$ -- solid line for model parameters \MI\ $\mathbf{P_1}$ and initial conditions $\mathbf{I_1}$ \eqref{P1} for $n = 5$.}

In Fig. \ref{ignatev4} shows a picture of the development of instability depending on the value of the wave number $ n $. In this figure, one can see that the increment / decrement of oscillations, firstly, inc\-reases with decreasing wavenumber, and, secondly, with decreasing wavenumber, the duration of the insta\-bility phase inc\-reases. Third, in Fig. \ref{ignatev4} it can be seen that the duration of the instability phase can increase only due to an earlier start of this process, the process is frozen at the same time, regardless of the value of $ n $. In addition, let us pay attention to the fact that according to the initial conditions \eqref{P1} at the moment of time $t = 0$ the phantom field is actually absent $\varphi (0) = 10^{- 5}$, while has the potential $\Phi (0) = 1$ as a canonical field. In \cite{STFI_20} it is specially noted that the phantom field is a stabilizing factor that counteracts the development of instability in the cosmological system. Indeed, when the initial conditions \eqref{P1} permute the cano\-nical and phantom fields, the growth rate of oscil\-la\-tions vanishes\footnote{We do not present the corresponding schedule due to its uselessness.}.
\Fig{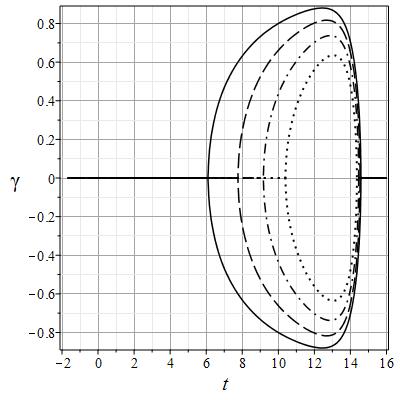}{8}{\label{ignatev4}Evolution of the increment / decrement of the increase / decrease of the oscillation amplitude $\gamma^\pm$ depending on the value of $n$ for the model parameters \MI\ $\mathbf{P_1}$ and the initial conditions $\mathbf{I_1}$ \eqref{P1} : solid line - $ n = 4 $, dashed line - $ n = 5 $, dash - dotted line - $ n = 6 $, dotted line - $ n = 7 $.}

Fourth, a comparison of the graphs in Fig. \ref{ignatev2} and \ref{ignatev3} show that the moment when the oscillation growth increment vanishes coincides with the moment of the minimum of the Hubble parameter.

Note further that the amplitude of perturbations at time $ t $ is determined by the expression
\begin{equation}\label{chi}
\chi(t)=\exp \left(\int\limits_{t_0}^t \gamma(t)dt\right),
\end{equation}
where $ t_0 $ is the initial instant of instability onset. Let $ t_1 $ be a finite moment of time, so that for $ t> t_1 $ $\ gamma (t_1) = 0$. Thus, during the development of instability, the disturbance amplitude is fixed at the value $\exp (\chi_infty)$:
\begin{equation}\label{chi8}
\chi_\infty=\exp \left(\int\limits_{t_0}^{t_1} \gamma(t)dt\right),
\end{equation}
Fig. \ref{ignatev5} shows the plots of $ \ln\chi (t) $ versus the wavenumber $ n $. The finite values of $\exp(\ln\chi) $ just give the finite values of the perturbation amplitude $\chi_\infty$ depending on the wave number. As you can see from the graphs, the value of the final amplitude changes in the range $\exp(\chi_\infty) = 7 \ div 340 $ when the wavenumber changes within $ [7,4] $. This growth of the disturbance amplitude occurs in the time interval $[t_0, t_1] \approx [6,14]$, ie, for $8$ units of time. Recall that we are working in the Planck system of units.

\Fig{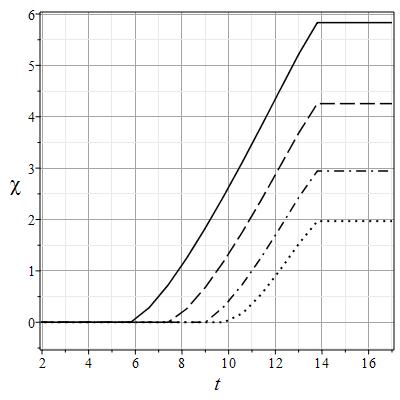}{8}{\label{ignatev5}Evolution of the logarithm of the perturbation amplitude $\ln\chi (t)$ depending on the value of $ n $ for the model parameters \MI\ $\mathbf{P_1}$ and the initial conditions $\mathbf{I_1}$ \eqref{P1}: solid line -- $ n = 4 $, dashed line - $ n = 5 $, dash - dotted line - $ n = 6 $, dotted line - $ n = 7 $.}
\subsection{A Scenario with the Hubble Parameter Sign Change}
We considered the previous case as an example, which can be used to demonstrate the main pro\-per\-ties of the development of instability in a cosmolo\-gical two-component system of scalar charged dege\-ne\-rate fermions. However, the given example has the disadvantage that the most interesting behavior of the system is found in the region of small values of the wave number, while the conditions for the applicability of the WKB approximation
On the other hand, \eqref{WKB} requires large values of $ n $. Therefore, the considered case is a formal example with an insufficient degree of reliability. So, for example, the condition \eqref{WKB} $ n^2 \gg e^4 $ for our case would look according to \eqref{P1} as $ 16 \gg 1 $ for $ n = 4 $.
In this regard, let us consider a scenario in which there is a change in the phases of compression and expansion. Note that this scenario is fairly typical for the \MI\  \cite{Ignat_GC21} model:
\begin{eqnarray}\label{P2}
\mathbf{P_2}=[[1,1,0.001,0.01],[1,1,0.5,0.001],0.1],\nonumber\\
\mathbf{I_2}=[1,0,0.0001,0.1,1].
\end{eqnarray}
This case differs from the one considered above by small values of scalar charges of fermions and small values of the initial Fermi momenta, but at the same time rather large potentials of the phantom field. In Fig. \ref{ignatev6} shows the evolution of the scale function $\xi(t) \equiv \ln a (t) $ for this case, corresponding to the sign change of the Hubble parameter at the time $ t \approx 0 $.
\Fig{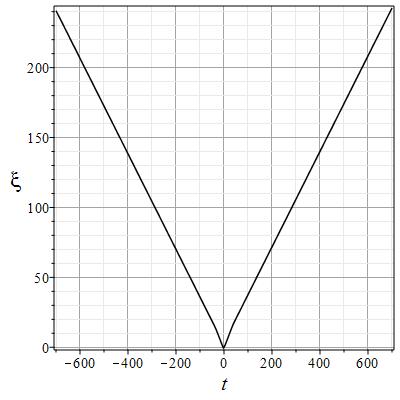}{8}{\label{ignatev6}Evolution of the scale function $\xi(t)$ for model parameters \MI  $\mathbf{P_2}$ and initial conditions $\mathbf{I_2}$ \eqref{P2}.}

Figure \ref{ignatev7} shows the evolution of the Hubble parameter for this case. It can be seen that the cosmological model makes a transition from the stage of inflationary contraction ($ H_{- \infty} \approx -0.34 $) to the stage of inflationary expansion ($ H_{+ \infty} = - H_{- \infty} \approx + 0.34 $ ). This transition occurs rather quickly and is accompanied by bursts of the Hubble parameter.

Fig. \ref{ignatev8} shows the evolution of the instability increment for the considered case and for sufficiently large values of the wave number $ n = 5 \div 100 $, under which the conditions of the WKB approximation \eqref{WKB} are satisfied with a large margin.

We see the emergence of two regions of instability: the first - at the stage of compression and the second - at the stage of expansion. In this case, the beginning of the first instability and the end of the second do not depend on the wave number - they coincide with the beginning of the first burst of the negative value of the Hubble parameter and with the end of the second burst of the positive value of the Hubble parameter.

\Fig{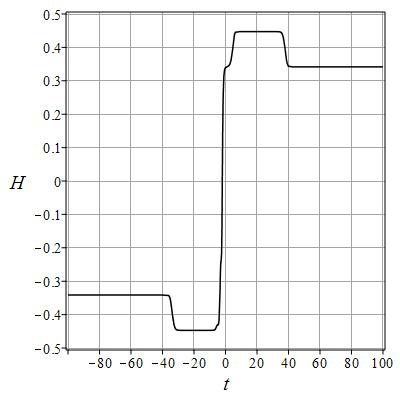}{8}{\label{ignatev7}Evolution of the Hubble parameter $H(t)$ for model parameters \MI\ $\mathbf{P_2}$ and initial conditions $\mathbf{I_2}$ \eqref{P2}.}
\Fig{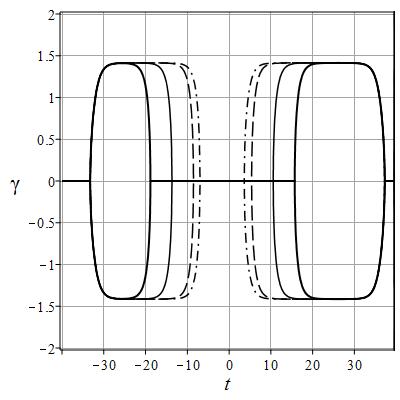}{8}{\label{ignatev8}Evolution of the increment / decrement of the increase / decrease of the oscillation amplitude $\gamma^\pm$ depending on the value of $n$ for the model parameters \MI\ $\mathbf{P_2}$ and the initial conditions $\mathbf{I_2}$ \eqref{P2} for wave numbers $ n = 5 $ - dash - dotted line, $ n = 10 $ - dashed line, $ n = 100 $ - solid line, $ n = 1000 $ - bold line.}

In this case, the beginning of the first instability and the end of the second do not depend on the wave number - they coincide with the beginning of the first burst of the negative value of the Hubble parameter and with the end of the second burst of the positive value of the Hubble parameter. The end of the first instability and the beginning of the second, with a decrease in the wavenumber, shift to the moment of the change of the phases of cosmological contraction and expansion. In this case, the duration of the first instability changes in the range of $13\div 24 $ and on average about $20$ time units, the second one changes in the range of $ 20\div 32$ and on average about $26$ time units. Thus, at the stage of compression, on average, $\chi_\infty \sim 29$, and at the stage of expansion $\chi_\infty \sim 38$, which corresponds to an increase in the amplitude of perturbations by 12 and 16 orders of magnitude (!), respectively.

Fig. \ref{ignatev12} shows the $\chi_\infty(n)$ dependency for the investigated case.
\Fig{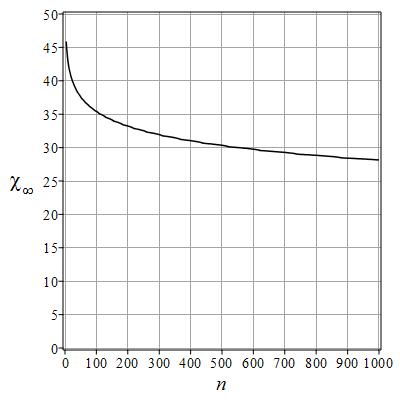}{8}{\label{ignatev12}Dependence of $\chi_\infty$ \eqref{chi8} on the wavenumber $n$ for the parameters $\mathbf{P_2}$ and the initial conditions $\mathbf{I_2}$ \eqref{P2}.}
This figure shows that although $\chi_\infty$ falls with an increase in the wavenumber, this fall is insignificant: with an increase in $n$ 1000 times, the value of $\chi_\infty$ falls less than 2 times, and gives for the amplitude perturbations $\exp(\chi_\infty)$ difference by 1 order of magnitude. Therefore, it can be argued that the me\-cha\-nism of the revealed gravitational - scalar instability is fundamentally different from the me\-cha\-nism of the standard gravitational instability of nonrelativistic matter, in which instability develops when the perturbation wavelength of a certain Jeans length is exceeded. In addition, due to the Jeans instability, perturbations grow slowly with time according to a power law, while in our case they grow exponentially.

To better understand the nature of gravitational - scalar instability, it is necessary to find out what happens to scalar fields and fermions in instability regions. Fig. \eqref{ignatev9} shows the evolution of the unperturbed potentials of the canonical and phantom scalar fields for the case under study. It is easy to see that the beginning and end of the instability phase coincide with the increase and decrease in the value of the potential of the canonical scalar field. The pause between two instability intervals roughly coincides with the region of the phantom field potential drop. On the other hand, as can be seen from the comparison of the graphs in Fig. \ref{ignatev7} -- \ref{ignatev9}, this pause coincides with the interval of the fast transition of the model from compression to expansion.
\Fig{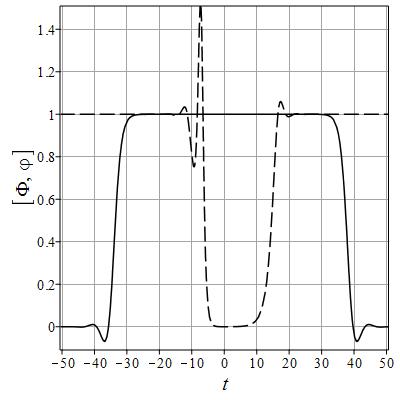}{8}{\label{ignatev9}Evolution of the potentials of the scalar doublet fields for the parameters of the model \MI\ $\mathbf{P_2}$ and the initial conditions $\mathbf{I_2}$ \eqref{P2}: $\Phi(t)$ is a solid line, $\varphi(t )$ - dashed line.}

It should be noted that those shown in Fig. \ref{ignatev9} plots show the process of generation of the canonical field $\Phi$ by phantom $\varphi$ due to the interaction between these components through the carrier of scalar charges - a system of degenerate fermions. Fig. \ref{ignatev10} shows the same picture, but on a smaller scale. It can be seen in this figure that the system starts from a stable singular point of the scalar doublet $\Phi = 0, \varphi = -1 $ and after the transient process ends up at another stable singular point $\Phi = 0, \varphi = 1$. Similar processes are observed in the \ MII\  model, which is based on a one-component system of degenerate doubly scalarly charged fermions.
\Fig{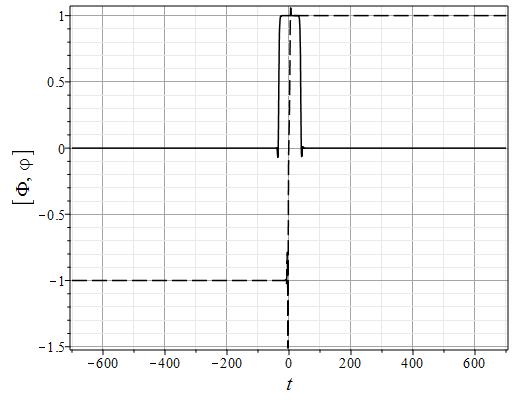}{8}{\label{ignatev10}Evolution of the potentials of the fields of the scalar doublet for the parameters of the  model \MI\ $\mathbf{P_2}$ model and the initial conditions $\mathbf{I_2}$ \eqref{P2} on a small scale: $\Phi (t)$ is a solid line, $\varphi (t)$ - dashed line.}
\Fig{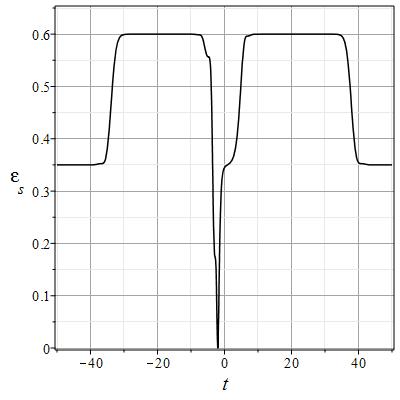}{8}{\label{ignatev11}Evolution of the energy density of a scalar doublet for the parameters of the  model  \MI\ $\mathbf{P_2} $ model and the initial conditions $\mathbf{I_2}$ \eqref {P2}.}

In Fig. \ref{ignatev11} shows the evolution of the total energy density of the scalar doublet (taking into account the cosmological constant). A narrow dip in this graph near the moment of the change in the compression and expansion phases (Fig. \ref{ignatev7}) is caused by a burst of negative energy of the phantom field (Fig. \ref{ignatev9}). It is at this moment, as calculations show, that a density burst occurs the energy of fermions, which, although it remains small in comparison with the energy of the scalar doublet, increases by a factor of $10^{10}$ at the instant of the burst.
This suggests that, although charged fermions make a very small contribution to the energy balance of the cosmological model, they play the role of a kind of catalyst that triggers the process of energy redistribution between the components of the scalar doublet.

\section*{Conclusion}
Summarizing the results of this work, we list its main results. \\[8pt]
$\bullet$\quad{\sl A numerical model of the evolution of gravitational - scalar perturbations in the cosmological model \MI, based on a two-component system of degenerate scalar charged fermions interacting through Higgs scalar fields, canonical and phantom, is constructed and tested.}
\\[8pt]
$\bullet$\quad{\sl Numerical simulation methods have confirmed the assumption of the appearance of gravitational - scalar instability of the uniform distribution of degenerate scalar charged fermions in the process of cosmological expansion. }
\\[8pt]
$\bullet$\quad{\sl Two cases of development of gravitational - scalar instability are presented and investigated: weak instability - for a cosmological model with a nonnegative Hubble and strong instability - for a cosmological model with an alternating Hubble parameter. It is shown that in the region of instability, disturbance oscillations have the character of growing stationary waves. }
\\[8pt]
$\bullet$\quad{\sl It is shown that a strong gravitational - scalar instability arises near the moment of a change in the phases of cosmological compression and expansion. In this case, two intervals of instability arise: in the compression phase and in the expansion phase. }
\\[8pt]
$\bullet$\quad{\sl The dependence of the growth rate of the perturbation amplitude on the wave number is investigated. It is shown that the growth rate of oscillations decreases with an increase in the wave number, but very weakly. It is shown that the intervals of instability coincide with the intervals of energy redistribution between the components of the scalar doublet and the intervals of growth of the energy density of fermions. }\\[8pt]
$\bullet$\quad{\sl Based on a comparison of the evolution of various components of the cosmological system and their perturbations, an assumption was made about the formation of gravitational - scalar instability due to the role of scalar charged fermions as a catalyst for the transfer of energy between the components of the scalar doublet. }

Thus, the results obtained show the promise of studying cosmological models based on systems of scalar charged particles.

\subsection*{Funding}
This paper has been supported by the Kazan Federal University Strategic Academic Leadership Program.
\end{document}